\documentclass[11pt]{article}

\usepackage[letterpaper,top=1in,bottom=1in,left=1.25in,right=1.25in,
            marginparwidth=1.05in,marginparsep=0.1in]{geometry}

\usepackage{microtype}
\usepackage{graphicx}
\usepackage{subcaption}
\usepackage{booktabs}   
\usepackage{caption}
\usepackage{amsmath}
\usepackage{amssymb}
\usepackage{mathtools}
\usepackage{amsthm}

\usepackage{natbib}     
\usepackage{hyperref}
\usepackage[capitalize,noabbrev]{cleveref}
\usepackage{xurl}       

\theoremstyle{plain}

\theoremstyle{definition}

\theoremstyle{remark}

\usepackage[disable,textsize=tiny]{todonotes}

\usepackage[final, commandnameprefix=ifneeded]{changes}
\definechangesauthor[name={PG}, color=blue]{PG}
\providecommand{\chcomment}{\comment}

\title{\texorpdfstring{\replaced[id=PG]{LLM-Aided Design for Manufacturing: A Multi-Agent System for Intent-Preserving Redesign of CAD for Improved Manufacturability}{LLM-Aided Design: Orchestrating Agents to Redesign CAD Parts for Improved Manufacturability}}{LLM-Aided Design for Manufacturing: A Multi-Agent System for Intent-Preserving Redesign of CAD for Improved Manufacturability}}

\author{
  Kojo Welbeck\,$^{1}$ \qquad Xiangyu Shi\,$^{2}$ \qquad Zahra Sadeghi\,$^{1}$ \\[0.3em]
  Qi Zhu\,$^{2}$ \qquad Ping Guo\,$^{1}$ \\[0.7em]
  \normalsize $^{1}$Department of Mechanical Engineering, Northwestern University, Evanston, USA \\
  \normalsize $^{2}$Department of Electrical and Computer Engineering, Northwestern University, Evanston, USA \\[0.4em]
  \normalsize Correspondence: \texttt{ping.guo@northwestern.edu}, \texttt{kojo.welbeck@northwestern.edu}
}
\date{Preprint. \today}

\begin{document}
\maketitle


\begin{abstract}
{We introduce autonomous, intent-preserving Design for Manufacturing (DFM) redesign of CAD parts: given an engineer's CAD model, the method returns a variant that is easier to manufacture without losing its design intent. Generating such a redesign in a single shot is unreliable, since CAD fidelity degrades as parts grow complex; we instead produce it as a sequence of individually verified design transitions. Our DFM-Redesign pipeline realizes this with two coupled agent subsystems driven by a pretrained multimodal LLM: a DFM Reviewer that inspects the current design and proposes one intent-preserving manufacturability improvement at a time, and a CAD Modifier that executes each proposal as an edit to the part's CadQuery program. The CAD Modifier closes a verification loop, compiling every candidate edit and visually checking it against the intended change from multi-view renderings, then re-generating or re-instructing until the edit is accepted or abandoned. Iterating review and verified modification compounds edits into parts more complex than one-shot generators reliably produce, preserves the original intent at each step, and requires no fine-tuning. On a 46-part benchmark scored by chamfer distance to reference geometries, the CAD Modifier reproduces target parts more accurately on average than chain-of-thought single agents given the same tools, and ablations isolate the contributions of the visual review loop and of captioning the design state before each edit. A centrifugal pump casing built from 32 chained transitions illustrates the complexity reachable by compounding verified edits. This is a preliminary report: evaluation of the full review-and-redesign loop, including manufacturability gain and an operational measure of intent preservation, is ongoing.}
\end{abstract}

\begin{figure}[tbp]
    \centering
    \includegraphics[width=\textwidth]{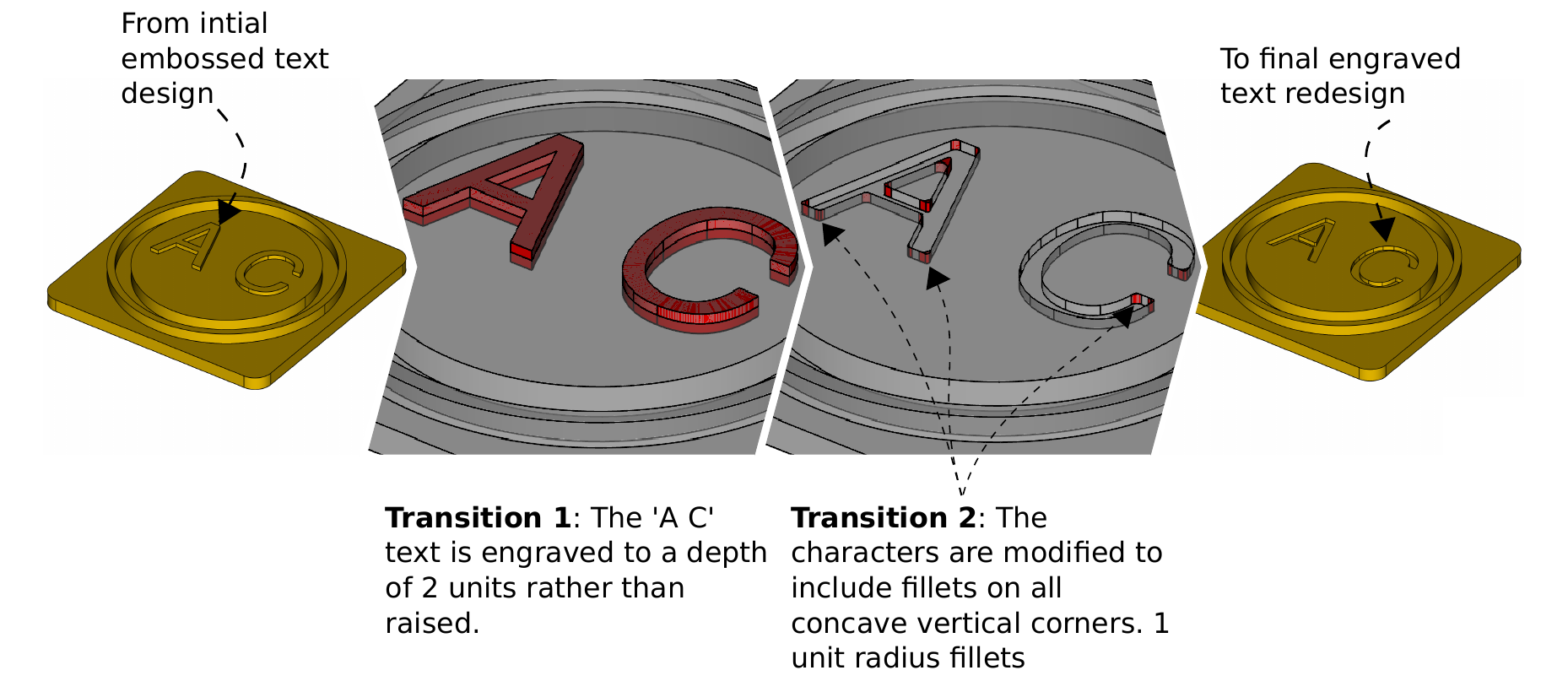}
    \caption{\textbf{Intent-preserving DFM redesign of an example part.} Starting from an initial design with raised (embossed) ``A~C'' lettering (left), the DFM-Redesign pipeline applies a sequence of individually verified design transitions: Transition~1 converts the raised text to text engraved $2$ units into the surface, and Transition~2 adds $1$-unit-radius fillets to the concave vertical corners of the characters. The final part (right) is more agreeable to CNC machining while retaining the original badge design intent.}
    \label{fig:teaser}
\end{figure}

\section{Introduction}
The possibility and the promise of automating the construction of 3D CAD models has driven the exploration of AI-based CAD generation in recent years. Automating CAD generation holds the potential for exponential productivity gains in multiple ways. It could accelerate the development of products through the ideation phase, through the virtual prototyping and analysis phase and  towards the detailing and generation of specifications for downstream manufacturing. \replaced[id=PG]{It also broadens access to CAD beyond trained design and manufacturing engineers. These gains, however, are limited by a persistent obstacle: the fidelity of AI-generated CAD degrades as parts grow more complex, which erodes the usefulness of generated models for downstream engineering reasoning.}{It can broaden access to the CAD endeavor of virtually creating parts and assemblies of parts without the presently necessary expertise of mechanical design and manufacturing engineers. And with more stakeholders and experimentalists capitalizing, AI-based CAD generation can unlock more creativity and diversity in the product space.}\chcomment[id=PG]{Cite evidence for this complexity-fidelity claim: e.g.\ DeepCAD's segment and command caps, and reported accuracy-vs-complexity from the X-to-CAD papers. One concrete number or plot makes the motivation much stronger.}

In this work, we aim to take a step beyond the X-to-CAD paradigm that describes much of the work in the space of conditional CAD Generation research. By that paradigm, generative models are trained to generate CAD by translating from one or more modalities. For example, GenCAD~\citep{alam2024gencad}, Img2CAD~\citep{chen2024img2cad} and CADDreamer~\citep{li2025caddreamer} present different diffusion-based pipelines to generate different CAD representations from single view images of parts to be translated. \replaced[id=PG]{Instead, we extend conditional, controllable CAD generation to a setting where an AI model, equipped with prior knowledge, pursues the higher-level reasoning goals of a mechanical design engineer.}{We take aim, instead, at extending conditional and controllable CAD generation into a paradigm of where AI-models, equipped with prior intelligence, can be set to task on any of the numerous concurrent, higher-level reasoning goals of a human mechanical design engineer.} We opt, therefore, to generate CAD designs conditioned on other CAD designs using existing powerful and multimodal large language models. Specifically, we demonstrate how a multi-agentic system can be orchestrated to review a CAD model started by a design engineer; to assess the manufacturability of the design in its current state; to make recommendations to improve the design and manufacturability of the part; and to change the design state and construction sequence of the CAD model. In the end, for a given input CAD design, \replaced[id=PG]{the system of agents returns a revised, improved design}{the system of agents return a revised improved design}.

\replaced[id=PG]{This fidelity barrier shapes our approach. Our key novelty is to recast CAD generation as a sequence of individually verified design transitions: rather than translate a whole part in one pass, we apply each edit to the current model and verify it before applying the next. Compounding verified transitions lets a general-purpose multimodal LLM build up complex parts without losing design intent, reaching designs that one-shot X-to-CAD methods cannot reliably produce, and it is what makes intent-preserving redesign for manufacturing feasible. \Cref{fig:teaser} shows this transition loop on an example part.}{Central to this work is a focus on accurately and reliable translating design intentions into the corresponding CAD artifacts. Thereafter, it becomes possible to steer CAD generation towards our objective of re-designing for manufacturing. However,  across related research, we observe the trend of decreasing fidelity to design intentions in CAD generation as the complexity of the parts increase. Consequently, there is a corresponding diminishing utility of said models to subsequent reasoning processes as parts become complex.  Aligned with the objective of modifying part designs, we generate CAD models by effecting design transitions singly and compounding sequentially and by verifying the updated model after each transition. This gives us a pathway towards complex part designs that would previously be unattainable by current methods.} 

The design for manufacturing know-how comes directly from a general purpose pre-trained model. It is sufficient for a proof of concept without the need to furnish quality training data (which is not yet available in appropriate quantities) and/or to finetune. Particularly, we use multimodal Gemini models, which allows for a visual inspection and interrogation of design parts when reviewing for manufacturability and when assessing the evolution of parts re-designed towards a desired end. We limit the manufacturability analysis to focus on CNC machining only.

\replaced[id=PG]{In summary, this work makes three contributions.}{All told, the key contributions laid out in this work is as follows:}
\replaced[id=PG]{\textbf{A transition-wise CAD generation method.} We build a part as a sequence of individually verified edits, instructed in natural language and checked through image-based feedback loops, reaching higher complexity than one-shot generation while preserving design intent.}{Towards high complexity CAD outputs, we introduce a transitions-wise CAD generation multi-agentic subsystem that is text-instructed through multi-turn changes in a CAD part model. The subsystem is tooled to prompt Gemini models to generate CadQuery scripts, and features image-based feedback loops to verify and improve accuracy of generated CAD modifications.}

\replaced[id=PG]{\textbf{The DFM-Redesign pipeline.} We embed this method in an autonomous, end-to-end multi-agent system that reviews a CAD part and redesigns it to be more agreeable to CNC machining while retaining the original intent, using a general-purpose multimodal LLM without fine-tuning.}{We incorporate the aforementioned CAD transitions generator into an autonomous end-to-end review and redesign system. Also built around image-based inspection and prompting Gemini models, the system is purposed to produce design alternatives that are more agreeable to CNC machining while retaining the intent of the original designs.}

\replaced[id=PG]{\textbf{Evaluation protocols and datasets.} We propose evaluation tests and heuristics for intent-preserving DFM redesign and release datasets to support comparable experimentation.}{The end-to-end system DFM review system and the CAD modification subsystem are evaluated using selections of CadQuery scripts. We propose different evaluation test and heuristics and provide respective datasets for similar experimentation.}\chcomment[id=PG]{Add a one-line forward pointer to the headline result here once the evaluation is run (e.g.\ ``On [N] parts, ...''), and cross-reference the Evaluation section.} 	

\section{Relevant Work}
The task of AI-based CAD review and redesign for manufacturability is, arguably, itself a frontier task in conditioned CAD Generation. It introduces the notion of CAD generation, not only conditioned on the current design state, but also on the intent to improve parts for manufacturability. \replaced[id=PG]{Encube~\citep{encube2025} describes a similar conception in broad strokes without a view into the implementation details, and AgentsCAD~\citep{george2026agentscad} applies multi-agent LLM reasoning to design for fused-deposition printing; we instead build and evaluate a concrete, intent-preserving, verified, training-free system for this task.}{Encube research describes a similar conception in broad strokes without a view into the implementation details.} In this section, we highlight some different tools, techniques and ideas that paved the way and which we borrow.

\subsection*{CAD as a Precise Geometric Language}
Behind the visual and oft manipulable virtual representation of CAD models is an (or multiple) equivalent language-based representation(s). While the major commercial CAD software abstract away the language-based representation(s) from the view of the designer, behind the scenes the textual information (as proprietary code,  serialized data structures, etc) encodes the command sequence and all the parametric geometric and topological details of the 3D parts and assemblies. The visual objects created are often interrogable and editable through a WIMP ("windows, icons, menus, pointer") and dialog-based graphical user interface. CadQuery, on the other hand, features increasingly in CAD related research because it is an increasingly mature open-source codebase that allows users to design parametric 3D models in Python code. We similarly base our work on the CadQuery library, despite the comparative lack of adequate parts and assemblies datasets. Most importantly, however, the corpus on which recent Gemini models were trained included CadQuery documentation and examples, the evidence of which propelled this work.  Lastly, a CAD part in CadQuery script can also be exported into different formats for interoperability between CAD software, or as images, which we take advantage of as well. \added[id=PG]{We therefore operate directly on editable CadQuery programs, a representation that both the LLM and a human designer can read, run, and modify.}

\subsection*{Using Textual Visual Equivalence}
The tandem of, and the equivalence of, the underlying text and the visual 3D geometry captures the sequential nature and the state-based nature of CAD, spanning two modalities, at least. The sequence-to-state or text-to-3D shape mapping for any given CAD part is generally non-injective as there are multiple command options, command sequences and/or exponentially many permutations to derive the same visual 3D prototype. 
Text-to-CAD research aims to train models to translate conversational descriptions of parts into either 3D shapes directly or into a command sequence format or script that, when verified and executed, produces the 3D geometries. As part of building training datasets that map descriptions to the corresponding CAD, these works typically employ large language models as part of an annotation pipeline where image captures of the CAD model are processed by captioning models to produce AI-generated textual descriptions. We, similarly, take advantage of this text-vision equivalence and incorporate a visual feedback loop and annotating step into the multi-agentic system. The AI-evolved CAD is repeatedly converted into images which are inspected for accuracy ideally until congruence with the intended part redesign. \added[id=PG]{Unlike text-to-CAD annotation pipelines that use images only to caption training data, we place multi-view renderings inside a closed verification loop at generation time.}

\subsection*{From Direct Generation to Piece-wise Generation}
The X-to-CAD results unsurprisingly show a decrease in accuracy with increasing complexity of the intended design, where complexity has been described heuristically by different measures: the number of faces in the part, the number of commands in the command sequence, etc.~\citep{wu2021deepcad,alam2024gencad,xu2024cadmllm}.\chcomment[id=PG]{Citations added. Still worth quoting a concrete accuracy-vs-complexity number from these works --- it is the empirical backbone of the motivation.} 

Typically, models trained for the X-to-CAD tasks will be limited by the complexity of models seen in its training dataset: the original source and derived datasets are static; they contain a limited range of CAD models; and, regardless of how complex a command sequence may be, it is finite in length. Some examples in the literature purposely put hard limits on the complexity of CAD models to be generated. DeepCAD~\citep{wu2021deepcad}, re-representing CAD as a fixed-sized vector encoding of the command sequence, pre-filters its own dataset to exclude parts with more than a threshold number of segments of a sketch, sketches for an extrusion or commands in a sequence.

By avoiding supervised fine-tuning, we avoid incorporating the limitations on any one or group of datasets. Further, towards parts with greater complexity,  we focus on generating CAD modeling transitions to existing CAD designs, instead of directly generating desired models in one go like other CAD generation systems. Consequently, increasingly more complex CAD models can be built up by chaining more feature transitions (commands) to an evolving command sequence and by extension an evolving design state of a part. While a guiding motivation is to produce human editable CAD, we demonstrate AI-editable CAD by this piecewise approach. The conditions are thus favorable for bidirectional human-AI collaboration. The piecewise approach used in this work also allows piece-wise review to theoretically maintain higher fidelity towards intended design targets.

\replaced[id=PG]{Sequential, designer-like construction is rare in AI-based CAD generation, and the few step-wise methods differ from ours in control and conditioning.}{The idea of sequential generations appropriately mimics the process of a mechanical designer using CAD software, but this is rare but not absent in the AI-based CAD generation literature.}
\replaced[id=PG]{BrepGen~\citep{xu2024brepgen} denoises a boundary representation face-, edge-, then vertex-wise, but it is only loosely conditioned and assembles a single fixed output rather than an editable, instruction-driven sequence. FlexCAD~\citep{zhang2025flexcad} edits an existing model by randomly regenerating a feature at a chosen level of the CAD hierarchy, so it is piecewise but the change it makes is uncontrolled. CAD-MLLM~\citep{xu2024cadmllm} instruction-tunes on a specially annotated command sequence to add or delete components, but it needs task-specific training data. In contrast, our transitions are goal-directed, individually verified against the intended change, and produced by a general-purpose multimodal LLM with no fine-tuning, which is what lets us compound controlled edits toward complex, intent-preserving redesigns. CAD-Llama~\citep{li2025cadllama} likewise adapts a general LLM to parametric CAD, but through domain-specific pretraining rather than at inference time.}{BRepGen uses a transformer-based diffusion model to generate CAD boundary representation models by denoising faces, then edges then vertices. Though step-wise, BrepGen is loosely conditioned, and the steps together generate a single completed boundary representation, not . FlexCAD modifies existing CAD by randomly regenerating a replacement feature at a selected level of the CAD model. This CAD generation, though piece-wise, is uncontrollable in the changes made. CAD-MLLM shows how models trained on its specially-annotated CAD command sequence can be instruction-tuned to add and delete specific components of the CAD by text instructions.}\chcomment[id=PG]{Confirm coverage of recent competitors (2024-2026 LLM/agentic-CAD and text-to-CAD). CAD-Llama is in our references but uncited; decide whether it and similar works belong here.}

\section{Methodology}

In this section, we describe the end-to-end {DFM-Redesign pipeline}, a multi-agent system that takes in a representation of the initial design state of a CAD part and returns that of the redesigned state of the part, for which an attempt is made to make machining easier, faster and/or cheaper. 
We describe the {DFM-Redesign pipeline} as a combination of two subsystems: the CAD Modifier and the DFM Reviewer, plus the ancillary infrastructure they depend on as preliminaries.

\begin{figure}[tbp]
    \centering
    \includegraphics[width=\textwidth]{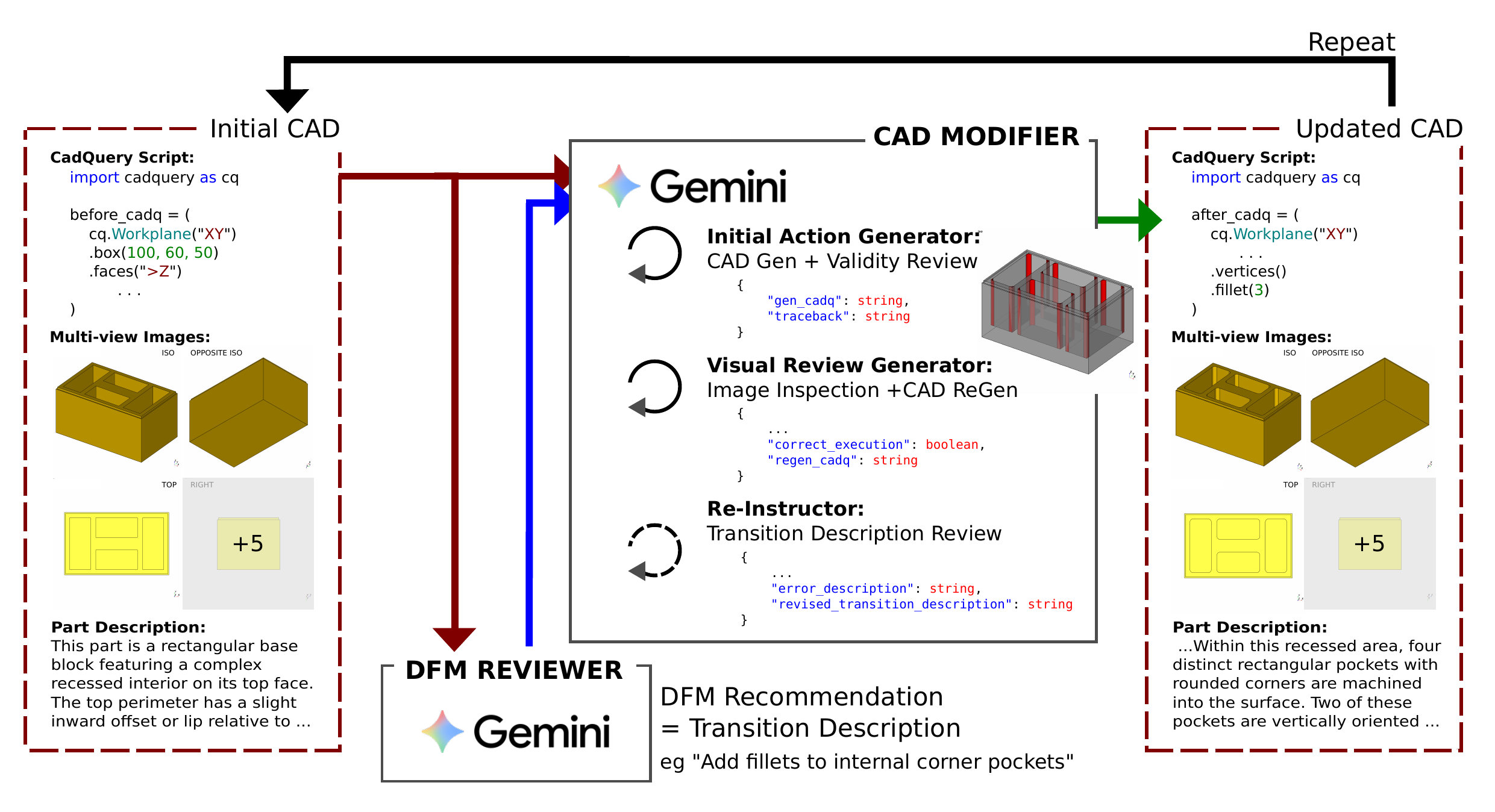}
    \caption{\textbf{The DFM-Redesign pipeline.} A DFM Reviewer (a multimodal Gemini agent) inspects the current design---its CadQuery script, multi-view renderings, and textual part description---and emits a single manufacturability recommendation as a natural-language transition description. The downstream CAD Modifier executes that transition through three agents: an \emph{Initial Action Generator} that produces a compile-verified ``after'' CadQuery script, a \emph{Visual Review Generator} that checks the rendered result against the intended change, and a \emph{Re-Instructor} that re-phrases the transition when visual validation repeatedly fails. The accepted ``after'' design becomes the input to the next review--modify cycle, compounding verified transitions until no further recommendation is warranted.}
    \label{fig:method}
\end{figure}

\subsection*{CadQuery and Multi-view Images}
CadQuery~\citep{cadquery} is the chosen modeling script for its intuitiveness and sufficient maturity, and because its use has been verified as compatible with recent Gemini models. 

It is self-contained and installs prepackaged with vtk visualization tools that allows saving screen shots of the part or assembly modeled by the script. So accordingly we render each CAD model in 8 preset perspectives, labeled with perspective names (Isometric, Front, Rear, Right, Left, Top, Bottom, Opposite Isometric) and with the trihedron displayed to disambiguate the respective images in the multi-view set.

CadQuery, furthermore, allows boolean operations between one or more models, which allows us to extract the differences between two states in the evolution of a part. By recoloring said extracted differences, we highlight and capture the transition features between two design states in a “transition” multi-view image set.

\subsection*{CAD Modifier}
The CAD Modifier is downstream of the DFM Reviewer and is the central workhorse of the end-to-end system. It could potentially be paired with differently-intentioned upstream subsystems to achieve other objectives beyond design for manufacturing.

As conceived, it requires information about the current (“before”) state of the design — namely, the CadQuery script of the command sequences that creates the design, a multi-view rendered image set and a textual description of the part — and instructions about or a description of the desired transition effected upon the design. At the end of its process, it returns the CadQuery script of the next evolution state of the CAD design. By sequentially processing through different transitions, an increasingly more complexly evolved part can be created. 

The CAD Modifier uses three agents in its execution:

The first is the \textbf{Initial Action Generator} that receives the current (or “before”) state of the CAD. It expects a high level text description of the shape and features of the part,  a multi-view image set to further its understanding of the geometry and topology of the part and its features, and the CadQuery script content to reason about the sequential and relational construction and to understand the parametric (dimensional) details of the part. From this context, the large language model processes the transition or “delta” description to generate an updated (or “after”) CadQuery script. This script corresponds to the original design stepped forward in state to include the instructed design change. The generated CadQuery script is executed in a testbed environment to verify successful compilation of the Python script. The Initial Action Generator is implemented as a chat client that chains together any preceding  history of unsuccessfully verified scripts and the traceback message from the respective compilation attempts as further context for any subsequent CAD generation attempts. A system parameter specifies how many attempts the agent has before abandoning the commanded transition.

The second agent, the \textbf{Visual Review Generator},  through a visual feedback loop, checks the recent compilation-verified “after” CadQuery script for congruence with the intended design change. If the validation check fails, the agent, also implemented as a chat client, continually attempts to refine the generated “after” script, considering the history of all previous attempts.

For inputs, the LLM expects the “before” and “delta” textual descriptions and the visual artifacts that went into and came out of the first agent, including the “after” and “transition” multi-view image sets derived from the generated and verified CadQuery model script. Until convergence or until the process is terminated early by a specified maximal number of retries, the agent determines if the generated “delta” JSON correctly executes the transition instruction by inspecting the multi-view images. In the event of a failed transition, the agent describes the error which, by the chat implement, informs the next regeneration and visual feedback assessment attempt by the agent. Alternatively, the Visual Review Generator accepts and returns the generated “after” JSON as correct, vis-\`a-vis the delta description, and then generates an updated caption (part description) describing the newly-revised design state of the part. The new CadQuery script and the new part description are directly forwarded through as outputs of the CAD Modifier and as a primer for any further design evolutions through the CAD Modifier.

Intermittently, after a fixed frequency of non-converging loops by the Visual Review Generator, the CAD Modifier calls on the \textbf{Re-Instructor} agent to re-phrase the transition description. It considers the same “before” multimodal artifacts, the current “delta” description and the latest attempt and renderings from the second agent to (1) describe the design transition of the latest attempt, (2) critique it against the intended design transition and (3) generate a new transition description that is, at the same time, faithful to the design intent but a different approach from the previous instruction.\chcomment[id=PG]{The Variant~A vs.\ Variant~B ablation in the Evaluation section finds this component to be a net negative and retires it from later runs. Reconcile: either keep it here as a described-then-ablated component, or drop it from the method.}

\subsection*{DFM Reviewer}
The DFM Reviewer is the higher-level task master. It contextualizes the system objective, of making designed parts to be machined easier to produce, to any given design. To wit, it translates the DFM objective into specific context-aware design transition recommendations to be executed as instructions by the downstream CAD Modifier. 

We prompt the language model to generate a single transition suggestion at a time based on the most recent design state of the CAD model, to ensure a tighter and better coupling to the evolving design. The alternative of generating multiple suggestions based on a previously time-stamped design state would often produce instructions that are soon out of context and  that conflict with a later evolved state of the CAD model as the transitions are singly effected out. On the other hand, a chaining of singly generated in-context recommendations singly and immediately implemented by the CAD Modifier leads to further refinements in the design, sometimes as multi-stage design corrections. 

Reliant on the “intelligence” acquired by the multi-modal language model during pre-training, the DFM Reviewer module processes the “before” CadQuery script by inspecting the multi-view image set of the current design state. It then returns a suggested change to the design along with a title and its reasoning for the design change. 

The large language model is prompted to think through various aspects of turning the design into a physical part by machining, so as to encourage a detailed analysis of feasibility and how the part will be made. These include questions about blank sizes, about enumerating the different operations, about part setups and reorientation, and about tool selection. Answers to these are not included in the reported output, but are fleshed out in the model’s “thinking” output.

We pay extra attention to further constraining the sorts of DFM suggestions recommended by the subsystem to, as much as possible, preserve the design intent of the modeled parts by providing a set of 3 DFM guidelines. Specifically, the guidelines prompt the LLM to (1) avoid sharp internal vertical corners; (2) favor chamfers instead of horizontal external convex fillets and (3) avoid raised text that has to be machined. The model interrogates the current design to ascertain if any of the guidelines are applicable in the current context and/or if any has been violated by the current design. Notably, we explicitly prompt the model to output only recommendations that are necessary, and none otherwise, as a signal to end the evolution process. 

\section{Demonstration}

\begin{figure}[tbp]
    \centering
    \includegraphics[width=0.68\textwidth]{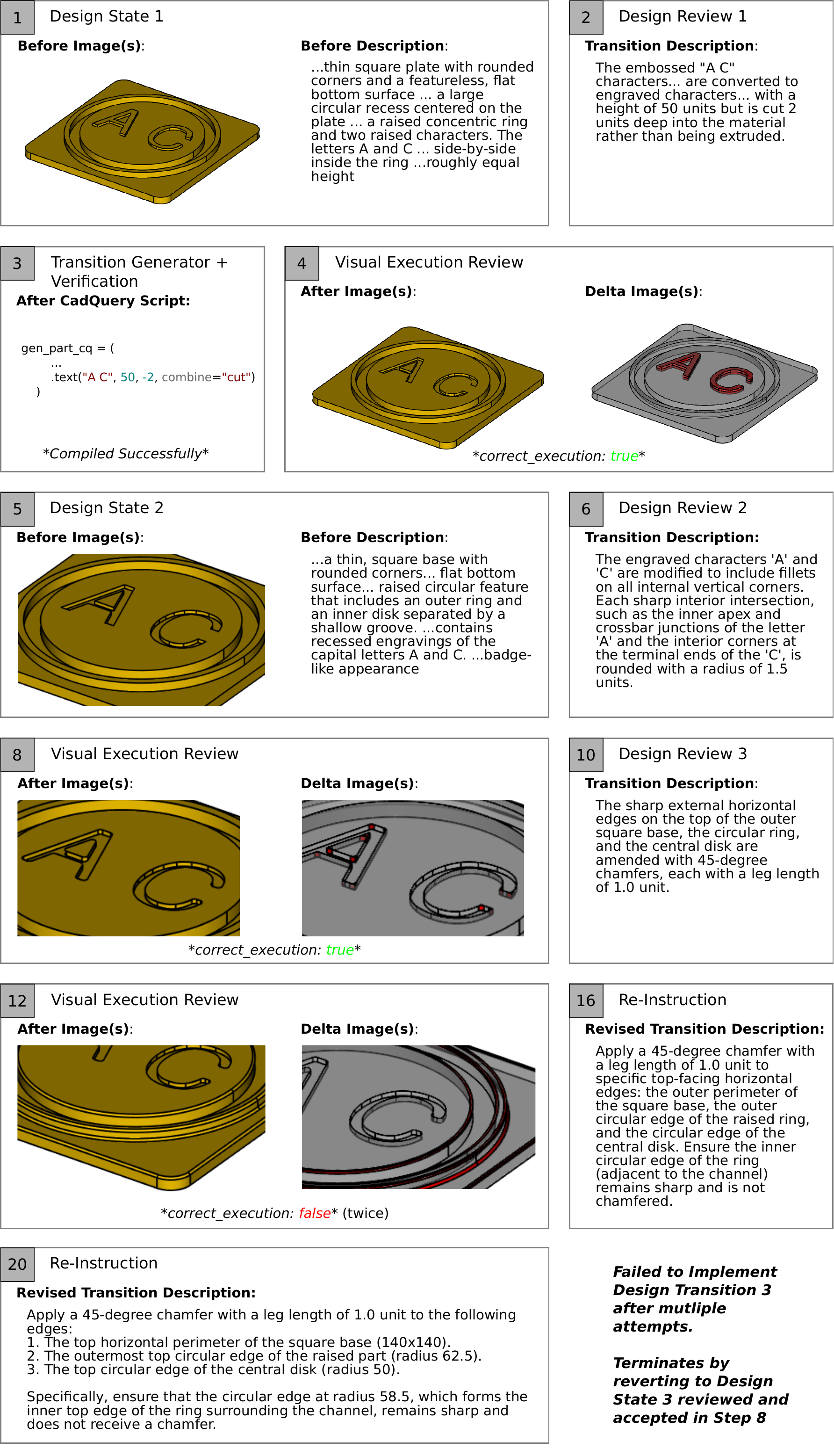}
    \caption{\textbf{Step-by-step redesign of the badge part.} Numbered panels trace the back-and-forth between the DFM Reviewer and the CAD Modifier. Transition~1 (embossed$\rightarrow$engraved text) is generated, compiler-verified, and visually accepted (\texttt{correct\_execution: true}). Transition~2 (fillets on internal letter corners) succeeds on the first attempt. The third recommendation (chamfer the top edges of the base, ring, and disk) is repeatedly compiler-valid but rejected by the Visual Review Generator---because the inner edge of the ring is also chamfered---triggering the Re-Instructor. After the maximum number of retries the system abandons the transition and reverts to the last accepted state (Design State~3) as the final design.}
    \label{fig:steps}
\end{figure}

\begin{figure}[tbp]
    \centering
    \includegraphics[width=0.95\textwidth]{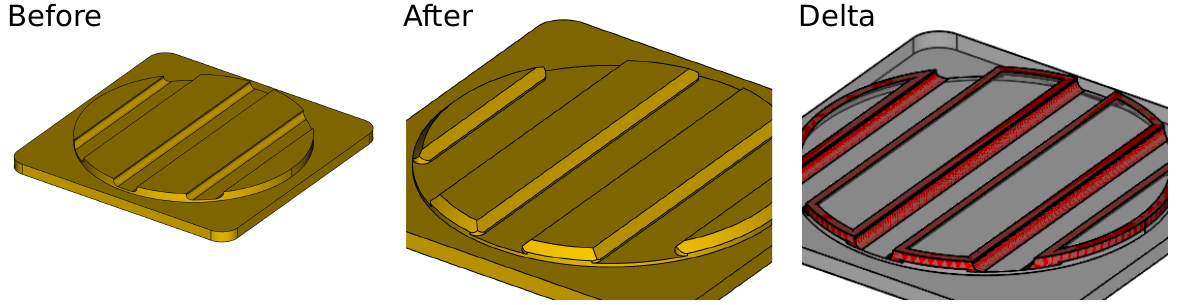}
    \caption{\textbf{Before, after, and delta for a slotted circular-boss part.} Isometric views of the original design (left), the redesigned part after processing by the DFM-Redesign pipeline (center), and the highlighted difference (right, in red) between the two states. The redesign replaces hard-to-machine horizontal convex fillets with $45^\circ$ chamfers and adds fillets to the internal vertical slot corners; the corresponding transition instructions and rationales produced by the DFM Reviewer are enumerated in \cref{tab:transitions}.}
    \label{fig:sample}
\end{figure}

In this section we show the results of the DFM redesign system anecdotally, before a systematic evaluation and benchmarking results in the section that follows. Here, we walk through how an original design sample evolves through the {DFM-Redesign pipeline}, through repeated back-and-forths between the DFM Reviewer and the CAD Modifier (\cref{fig:steps}). Also, we show a summary view of another sample to specifically highlight the behind-the-scenes reasoning of the DFM Reviewer (\cref{fig:sample,tab:transitions}). Then finally, we show a spread of more samples processed for DFM along with the transition titles.

The process flow (\cref{fig:steps}) begins with an initial design (Design State 1) of a badge-like part captured by its CadQuery script, a resulting multiview image set and an LLM-generated part description.
The DFM Reviewer processes the multimodal inputs to output a recommendation for redesign, namely to replace the embossed lettering with an engraved version. Then, the design transition is executed in Step 3 by the Initial Action Generator where the generated CadQuery script is compiler-verified, and then visually validated and accepted in Step 4. 
In Step~6, the DFM Reviewer processes the multimodal inputs of the just-updated design (Design State 2), recommending, this time, to add fillets to all the internal corners of the engraved letterings. This change is implemented correctly at the first attempt by the Initial Action Generator, and is thus accepted by the Visual Review Generator. Subsequently, a third pass of the evolving CAD through DFM Reviewer then CAD Modifier terminates unsuccessfully. The DFM recommendation to chamfer the sharp top edges of the base square, the outer ring and the central cylinder are repeatedly incorrectly executed as determined by the Visual Review Generator, though compiler-verified. The Re-Instructor is called on periodically to re-work the particular DFM suggestion. Successive attempts are again inspected visually and judged to be erroneous, particularly because the inside edge of the ring was also chamfered. After a maximal number of retries,  the multi-agent system reverts back to the latest accepted design (Design State 3) as the final design state.

We begin to see a couple likely pitfalls. First, the third and final DFM Recommendation was arguably unnecessary, adding extra production steps to the part. The unsuccessful execution is beneficial in this case. Secondly, the Visual Review Generator does not always correctly assess the attempted execution of a redesign, as it should have accepted the initial attempt during the third pass. By rejecting the first few attempts, and calling on the Re-Instructor to clarify intent, the intent was changed to exclude the inner edge of the ring.


Alongside the transition descriptions, the DFM Reviewer provides a title and a summary reasoning for each recommendation. \Cref{fig:sample} shows isometric images of a slotted circular-boss sample part that has been processed by the {DFM-Redesign pipeline}, including a highlighting of the difference (or ``delta'') between the initial and final design states. \Cref{tab:transitions} enumerates the transition instructions and accompanying rationale produced in-process by the three successive calls to the DFM Reviewer.

\begin{table}[tbp]
    \centering
    \caption{\textbf{Transition instructions and rationales for the part in \cref{fig:sample}.} Each column is one design transition produced by the DFM Reviewer, comprising a short title, the natural-language transition description passed to the CAD Modifier, and the manufacturability rationale generated alongside it.}
    \label{tab:transitions}
    \footnotesize
    \setlength{\tabcolsep}{3pt}
    \begin{tabular}{r| p{0.25\textwidth} | p{0.25\textwidth} | p{0.25\textwidth}}
        & \textbf{Transition 1}: & \textbf{Transition 2:} & \textbf{Transition 3:} \\
         \textbf{title} & Replace Horizontal Convex Fillets with Chamfers
         & Add Fillets to Internal Vertical Slot Corners & Replace Horizontal Fillets with Chamfers\\
         \textbf{description:} & The horizontal convex edges located at the top of the three rectangular slots, which are oriented parallel to the Y-axis, are modified from 3.0-unit radius fillets to 45-degree chamfers with a leg length of 3.0 units. These chamfers define the transition between the top surface of the circular boss and the vertical walls of the slots.
         & The internal vertical corners of the three rectangular slots, which are 15 units wide and 125 units long, are updated to include fillets. Each of the four vertical edges in each of the three slots is modified with a fillet radius of 3.0 units.
         & The external horizontal fillets located on the top curved edges of the four segments of the central boss are replaced with 45-degree chamfers of 3.0 units. This change applies to the outer perimeter of the boss segments where the top face meets the cylindrical side wall, excluding the straight edges of the slots which are already chamfered.\\
         \textbf{rationale:} & Horizontal convex fillets require specialized ball-nosed end mills and complex 3D tool paths to machine on a 3-axis mill. Replacing these fillets with 45-degree chamfers allows the part to be manufactured using standard chamfering tools and simpler 2.5D machining strategies, which significantly reduces machining time and production costs while still achieving the design intent of removing sharp edges.
         & These vertical corners are currently modeled as sharp 90-degree internal angles. Since machining is typically done with rotating cylindrical tools like end mills, a sharp internal vertical corner is impossible to achieve without expensive secondary operations like wire EDM. Adding fillets allows the slots to be machined efficiently with a standard end mill in a single setup, reducing manufacturing time and cost while also reducing stress concentrations in those corners.
         &Horizontal external convex fillets on curved paths require specialized ball-nosed end-mills and complex 3D toolpaths to machine the rounded profile. Replacing these fillets with 45-degree chamfers simplifies the manufacturing process by allowing the use of standard chamfering tools or basic milling operations with simpler 2D toolpaths, which significantly reduces machining time and cost while still eliminating sharp edges and reducing stress concentrations.
    \end{tabular}
\end{table}

\section{Evaluation}
\chcomment[id=PG]{This section is drafted from Kojo's write-up. It covers the CAD Modifier module only; the end-to-end pipeline axes are still an outline at the end of the section.}

\added[id=PG]{Before evaluating the end-to-end {DFM-Redesign pipeline}, we analyze the CAD Modifier module within it. This module holds the ability to make live edits to CAD parts, and so is at the heart of the entire system. We investigate and discuss the efficacy of the CAD Modifier through anecdotal demonstrations of creating complex CAD models; by evaluating the re-creation of reference parts using a chamfer distance metric; and through an ablation study on the gains or detractions of different subcomponents tested through the same sequences of transitions. Alongside the ablation study, we benchmark the system of orchestrated LLM calls against chain-of-thought-prompted single agents that are tooled with the same capabilities to function call as is necessary.}

\subsection*{Complex Part Demonstration}

\begin{figure}[tbp]
    \centering
    \includegraphics[width=\textwidth]{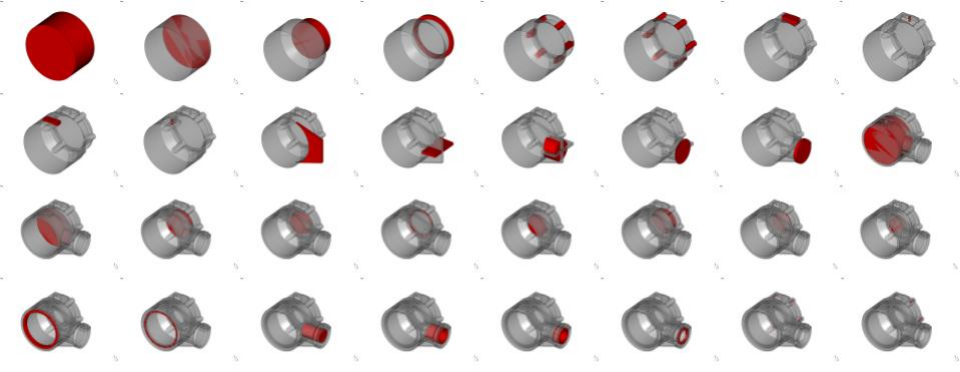}
    \caption{\textbf{Building a centrifugal pump casing over 32 chained transitions.} The most complex part generated by the CAD Modifier to date. Each panel is one verified design transition applied to the state produced by the previous one; the reference geometry was downloaded from GrabCAD and the transition descriptions were produced by a manually-supervised LLM annotation pipeline.}
    \label{fig:pump_build}
\end{figure}

\added[id=PG]{\Cref{fig:pump_build} demonstrates the process of building up the most complex part that has yet been generated by the CAD Modifier. Chaining together 32 transition descriptions sequentially, translated into corresponding CAD features, eventually produces what looks like a centrifugal pump casing. The original reference geometry was downloaded from GrabCAD. The descriptions were determined through a manually-guided and manually-supervised use of an LLM-based annotation pipeline. The exact ordering of CAD transitions was pre-determined manually and based on a familiarity with the CadQuery Python module. Images of the incremental changes were collected and fed to Gemini for per-transition captioning. The captions were reviewed and reworked as needed to be complete, understandable and unambiguous.}\chcomment[id=PG]{Source text read ``\ldots{}a centrifugal pump casing (Figure B)'', but Figure~B is the two-run variability figure, not the pump casing. Parenthetical dropped --- confirm with Kojo.}

\subsection*{Variability in Part Generation}

\begin{figure}[tbp]
    \centering
    \includegraphics[width=0.92\textwidth]{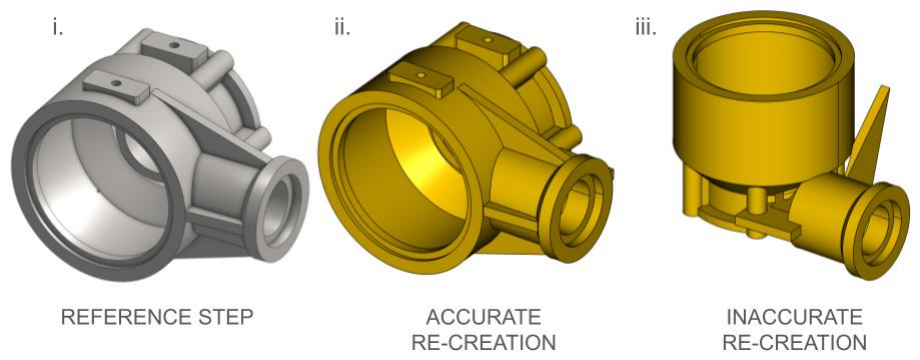}
    \caption{\textbf{Run-to-run variability under an identical transition sequence.} (i) The reference STEP geometry, against (ii) an accurate and (iii) an inaccurate re-creation, each produced by running the same 32 transition descriptions. In run (iii) the visual review loop accepted an incorrectly oriented base cylinder, after which features described relative to that base---the taper ribbing among them---were mis-applied.}
    \label{fig:variability}
\end{figure}

\added[id=PG]{The probabilistic nature of large language models shines through in how transitions build atop one another. Specifically, the same transition description or sequence of transition descriptions can yield different CAD changes, and so different final CAD parts. \Cref{fig:variability} shows two runs of attempting to re-create a reference part by incrementally effecting out the same 32 transition descriptions.}

\added[id=PG]{The most obvious source of variability is in the gating decision within the visual review loop on whether a successful change to the CAD part was consistent with the transition description instruction. Judging an incorrect change as accurate means continuing to build atop an erroneous foundation, and subsequent transition descriptions, which may include a mix of new features described in absolute and/or relative terms, may now be out-of-context. Other sources of variability include how the current state of the CAD design is interpreted from the combination of different inputs, and, as well, how the transition description is interpreted and contextualized.}

\added[id=PG]{During the inaccurate build of the pump casing, the review loop did not correctly orient the base cylinder and a number of subsequent features described in relation to that base were mis-applied; the taper ribbing, for example.}

\subsection*{Evaluation Dataset and the Chamfer Distance Metric}

\added[id=PG]{A slate of 46 other parts of differing sizes and complexity (read, number of transitions) were prepared into an evaluation dataset. The dataset featured the various image sets, STEP files and the manually-reviewed transition descriptions. The STEP files of the reference geometries enable a quantitative assessment of the CAD Modifier's accuracy in re-creating target geometries, via the chamfer distance computation upon point clouds of part pairs. This metric is adopted from and shared by other CAD generation efforts. In this work, the descriptions were specifically curated to start each build in the same orientation as the target geometry, to avoid such out-of-context continuations as seen in \cref{fig:variability}. We, however, relax this requirement in the chamfer distance assessments and employ a combination of global and local registration methods to align the point cloud of the generated CAD model to that of the target. An additional normalizing pre-processing step of scaling the point clouds according to the size of the reference parts allows for summarizing across the 46 parts.}\chcomment[id=PG]{Two edits to Kojo's sentence: ``qualitative assessment'' $\rightarrow$ ``quantitative'' (chamfer distance is a numeric metric), and the source pointed both this figure and the pump-casing figure at ``Figure~A''; the pump-casing build-up is \cref{fig:pump_build}; the complexity-spread figure it also pointed at has not been produced and is deferred. The source also cited ``Fig~B~iii'' --- panel labels not yet fixed.}

\subsection*{Ablation Study}

\added[id=PG]{In an ablation study for the optimal makeup of the CAD Modifier module, we investigated 6 variants on the same evaluation dataset, the summary statistics of which are shown in \cref{tab:ablation}. We expect the probabilistic nature of the generated CAD to account for some of the per-part differences generated by the respective variants. In addition to the raw averages and standard deviations, we discard some outliers for a more sanitized reporting of the statistics, and illustrate, along the way, examples of some errors observed.}

\begin{table}[tbp]
    \centering
    \caption{\textbf{Ablation summary over the 46-part evaluation set.} Chamfer distance statistics for the six CAD Modifier variants, reported over all parts (raw) and with the listed per-variant anomalies discarded. Best value in each column is bold; lower is better.}
    \label{tab:ablation}
            \small
    \setlength{\tabcolsep}{4pt}
    \begin{tabular}{lccccccc}
        \toprule
        & \multicolumn{3}{c}{raw} & & \multicolumn{3}{c}{without anomalies} \\
        \cmidrule(lr){2-4}\cmidrule(lr){6-8}
        & mean & median & std & anomalies & mean & median & std \\
        \midrule
        Variant A & 5.739 & 2.141 & 10.999 & [0, 30, 15, 24] & 2.943 & 1.700 & 3.673 \\
        Variant B & \textbf{3.496} & 1.567 & \textbf{9.183} & [28] & 2.181 & 1.492 & 2.219 \\
        Variant C & 4.474 & 2.047 & 9.482 & [28, 38] & 2.845 & 1.853 & 2.731 \\
        Variant D & 11.495 & 2.578 & 29.746 & [27, 18, 1] & 4.087 & 2.492 & 5.819 \\
        Variant E & 8.986 & 1.936 & 32.841 & [28, 0] & 3.329 & 1.775 & 4.286 \\
        Variant F & 3.634 & \textbf{1.398} & 10.030 & [28, 4] & \textbf{1.846} & \textbf{1.389} & \textbf{1.922} \\
        \bottomrule
    \end{tabular}
\end{table}

\todo[inline]{\cref{tab:ablation}: add a description column so each variant's makeup is readable in the same view (Kojo's note).}

\paragraph{Variant A vs.\ Variant B --- effect of changing the transition descriptions periodically.}

\added[id=PG]{Variant A features the full complement of subcomponents: the visual review loop, the periodic revision of the transition description that instructs the change to be implemented in CAD (the Re-Instructor), and an image captioning pre-processing step before the CadQuery generation. Variant B does not include the periodic revision of the transition descriptions.}

\added[id=PG]{Where Variant A reworks the instructing transition description at the 3rd and 5th rejection from the visual review, Variant B makes no description revisions. Incidentally, this revision is not triggered across all parts, so we highlight only the part runs that were affected by this subcomponent in \cref{tab:reinstructor}. We interpret the triggered samples as the harder samples that required more than 3 attempts at least once in the sequence of transitions.}

\begin{table}[tbp]
    \centering
    \caption{\textbf{Re-Instructor ablation, split by whether the description revision was triggered.} Chamfer distance statistics for Variant~A (with periodic transition-description revision) against Variant~B (without), separated into the 12 runs in which the revision fired and the 34 in which it did not. ``Lower Variant~B incidences'' counts the parts on which Variant~B achieved the smaller chamfer distance.}
    \label{tab:reinstructor}
            \small
    \setlength{\tabcolsep}{4pt}
    \begin{tabular}{lcccccccc}
        \toprule
        & \multicolumn{2}{c}{Mean} & \multicolumn{2}{c}{Median} & \multicolumn{2}{c}{Std.\ dev.} & & Lower B \\
        \cmidrule(lr){2-3}\cmidrule(lr){4-5}\cmidrule(lr){6-7}
        & A & B & A & B & A & B & Count & incid. \\
        \midrule
        Triggered & 6.229 & \textbf{2.411} & 3.576 & \textbf{1.386} & 7.552 & \textbf{2.758} & 12 & 8 \\
        Non-triggered & 5.566 & \textbf{3.879} & 1.700 & \textbf{1.674} & 12.076 & \textbf{10.577} & 34 & 20 \\
        \midrule
        Total & & & & & & & 46 & 28 \\
        \bottomrule
    \end{tabular}
\end{table}

\added[id=PG]{For both the 12 samples that triggered a revision of the transition descriptions through Variant A and the 34 samples that did not, we observed lower chamfer distance averages from Variant B, and at a higher rate of 28 of 46 samples. A deeper analysis of the changes made during the Variant A run revealed only a single description change, for one of the 12 parts, that ambiguated the meaning of the intended transition. That is to say, a demeriting effect of the periodic changing of the transition description is not straightforward to substantiate. We interpret the measured difference as having adversely affected or diluted the LLM's context prompt and state --- that is, additional information and responsibilities that limit its performance. Consequently, we omit the transition description revisions in subsequent runs and variants of the CAD Modifier, and attempt to review and sanitize the instructing transition descriptions in further upstream modules, when the transitions are first conceived in their proper context.}\chcomment[id=PG]{This finding retires the Re-Instructor, but the Methodology section still presents it as one of the CAD Modifier's three agents and the Demonstration narrates it firing. Decide with Kojo whether the Methodology keeps it as a described-then-ablated component or drops it.}

\paragraph{Variant B vs.\ Variant C --- effect of not including ``before'' descriptions.}

\begin{figure}[tbp]
    \centering
    \includegraphics[width=0.92\textwidth]{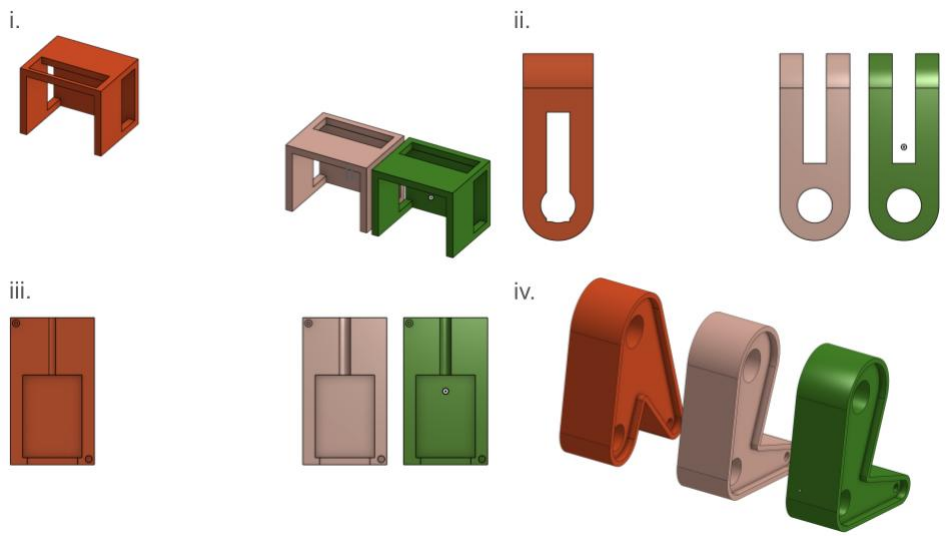}
    \caption{\textbf{Where the pre-transition caption helps.} Parts on which Variant~B (with the captioning step) tracks the reference geometry more closely than Variant~C (without it).}
    \label{fig:captioning}
\end{figure}

\added[id=PG]{Variant C is a spin-off of Variant B in which we omit the subcomponent that captions images of the current state of the CAD model before the CadQuery changes for a new transition are generated. In so doing, we assess the efficacy of generating the correct CadQuery script changes from just the current CadQuery and its image rendering, and whether the additional part description is a net advantage or disadvantage. The results, captured in \cref{tab:captioning}, show that more often than not Variant B produces lower chamfer distances. We interpret the net merit of including the pre-transition captions as furnishing more contextual (and summarized) understanding to the CadQuery script, the multi-view images and the transition descriptions. \Cref{fig:captioning} illustrates a few examples where the additional context guided the CAD generation towards a more accurate part compared to that generated by Variant C.}

\begin{table}[tbp]
    \centering
    \caption{\textbf{Effect of the pre-transition captioning step.} Per-part win counts and mean chamfer distance differences between Variant~B (with captioning) and Variant~C (without), including the subsets where the margin exceeds 1 and 4 units.}
    \label{tab:captioning}
    \begin{tabular}{lcc}
        \toprule
        & Count & Avg.\ difference \\
        \midrule
        Lower Variant B incidences & 31 & 1.712 \\
        Lower Variant C incidences & 14 & $-0.893$ \\
        Lower B by 1+ & 9 & 5.376 \\
        Lower C by 1+ & 3 & $-3.179$ \\
        Lower B by 4+ & 6 & 7.017 \\
        Lower C by 4+ & 1 & $-6.600$ \\
        \bottomrule
    \end{tabular}
\end{table}

\todo[inline]{\cref{tab:captioning}: Kojo suggests replacing this with a histogram. Note the win counts sum to 45, not 46 --- one part unaccounted for.}

\paragraph{Variant B vs.\ Variant D --- effect of the visual review loop.}

\begin{figure}[tbp]
    \centering
    \includegraphics[width=0.92\textwidth]{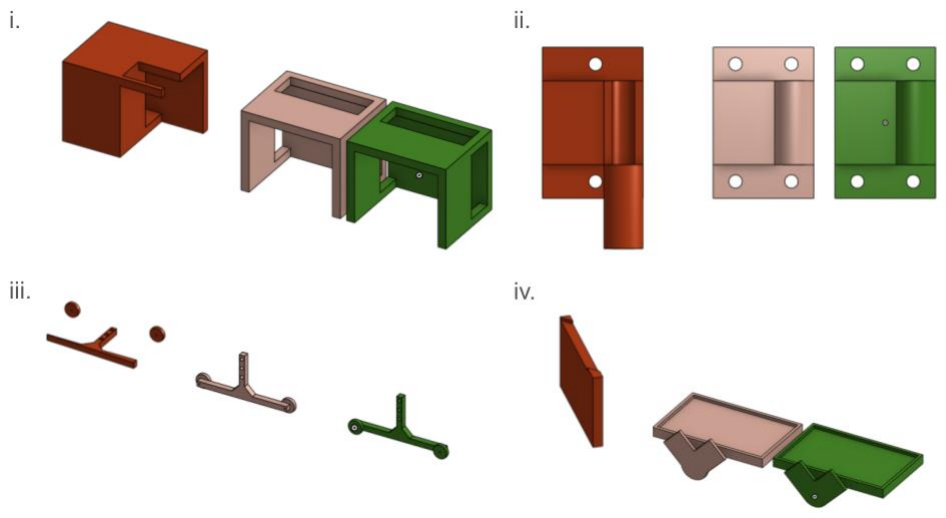}
    \caption{\textbf{Failure modes without the visual review loop.} With no gating agent (Variant~D), positioning and orientation errors go unchecked and compound across transitions, leaving final parts that diverge sharply from the design target.}
    \label{fig:visualreview}
\end{figure}

\added[id=PG]{Variant B compared to Variant D shows the largest effect of any component in this ablation. Removing the visual review loop raises the raw mean chamfer distance from $3.496$ to $11.495$, and from $2.181$ to $4.087$ with anomalies excluded --- a wider margin than any other pairwise comparison here. Without the loop there are larger and more frequent spikes in inaccuracy as measured by chamfer distance. The adjudicating agent within the review serves to prevent updates that are inconsistent with transition descriptions, and to inform the upstream CadQuery-generating agent where corrections are required. While we observe the porosity of this gating function in an earlier discussion on the sources of variability, we note now that the evidence shows that it is more helpful than it is a hindrance. \Cref{fig:visualreview} illustrates examples where, absent the visual review loop, the module is prone to more positioning and orientation errors. Without the built-in early termination logic of the visual review loop, errors compound and the final versions look very different from the intended design targets.}

\paragraph{Variant B vs.\ Variant E --- effect of including the history of previous transitions.}

\added[id=PG]{Variant E investigates whether an enumerated history of previous transition descriptions improves the generated CadQuery. The summary statistics say otherwise, but similar to the Variant A runs, the claim is not so straightforward to substantiate. The majority (35) show almost identical chamfer distance ($<1$) between the Variant B and E generations (\cref{tab:history}). Furthermore, the 7 samples that did not fare well in the Variant E runs have an outsized influence and skew the averages sizably.}

\begin{table}[tbp]
    \centering
    \caption{\textbf{Effect of supplying the history of previous transitions.} Per-part win counts and mean chamfer distance differences between Variant~B (no history) and Variant~E (with an enumerated transition history).}
    \label{tab:history}
    \begin{tabular}{lcc}
        \toprule
        & Count & Avg.\ difference \\
        \midrule
        Lower Variant B incidences & 29 & 8.939 \\
        Lower Variant E incidences & 17 & $-0.246$ \\
        Lower B by 1+ & 9 & 28.453 \\
        Lower E by 1+ & 2 & $-2.143$ \\
        Lower B by 4+ & 7 & 36.095 \\
        Lower E by 4+ & 0 & 0 \\
        \bottomrule
    \end{tabular}
\end{table}

\todo[inline]{\cref{tab:history}: Kojo suggests replacing this with a histogram.}

\paragraph{Variant B vs.\ Variant F --- effect of the choice of Gemini model.}

\added[id=PG]{The sixth and final variant, F, uses the newer Gemini 3.5 Flash model in place of the Gemini 3 Flash Preview model. Across almost all parts in the run, the measured chamfer distances are lower for Variant F. A finer inspection reveals Variant F made fewer compilation errors, required fewer retries through the visual review loop and, consequently, took less time across the 46 parts. See \cref{tab:model}.}\chcomment[id=PG]{Source text said ``Variant E'' three times in this paragraph where the comparison and Table~E are both Variant~F; corrected to F.}

\begin{table}[tbp]
    \centering
    \caption{\textbf{Effect of the underlying Gemini model.} Variant~B (Gemini 3 Flash Preview) against Variant~F (Gemini 3.5 Flash) over the 46-part evaluation set. Best value in each row is bold.}
    \label{tab:model}
    \begin{tabular}{lcc}
        \toprule
        & Variant B & Variant F \\
        \midrule
        Lower CD instances & 19 & \textbf{27} \\
        Completed transitions & 219/221 & \textbf{220/221} \\
        Failed reviews & 67/285 & \textbf{22/242} \\
        Failed verifications & 87/373 & \textbf{38/281} \\
        Average duration & 0:11:59 & \textbf{0:06:07} \\
        \bottomrule
    \end{tabular}
\end{table}

\subsection*{Benchmarking}

\added[id=PG]{Alongside assessing the components of the CAD Modifier through the 6 variants, we evaluated the same dataset on single agent systems that could themselves, through the function calling mechanism, pick and choose when (or whether) to (1) verify the successful compilation of the generated CadQuery script, or (2) render the ``before'', ``after'' or ``transition-highlighted'' multi-view images of the CadQuery script. In a total of 4 experiments we, once again, compare between the Gemini 3 Flash Preview model and the Gemini 3.5 model. On a second axis, we contrast between providing all transitions as one lumped input sequence and providing the transitions singly and between sequential attempts at generating the target geometries incrementally. \Cref{tab:benchmark} shows the summary statistics, including a comparison to Variants B and F from the ablation study.}

\begin{table}[tbp]
    \centering
    \caption{\textbf{Orchestrated CAD Modifier against chain-of-thought single-agent benchmarks.} Chamfer distance statistics over the same 46-part evaluation set, with Variants~B and~F carried over from \cref{tab:ablation} for comparison. Best value in each column is bold; lower is better.}
    \label{tab:benchmark}
            \small
    \setlength{\tabcolsep}{4pt}
    \begin{tabular}{lccccccc}
        \toprule
        & \multicolumn{3}{c}{raw} & & \multicolumn{3}{c}{without anomalies} \\
        \cmidrule(lr){2-4}\cmidrule(lr){6-8}
        & mean & median & std & anomalies & mean & median & std \\
        \midrule
        Variant B & \textbf{3.496} & 1.567 & \textbf{9.183} & [28] & 2.181 & 1.492 & 2.219 \\
        Variant F & 3.634 & 1.398 & 10.030 & [28, 4] & \textbf{1.846} & 1.389 & \textbf{1.922} \\
        \midrule
        Benchmark 1 & 6.812 & \textbf{1.384} & 22.135 & [27, 41] & 2.696 & \textbf{1.355} & 4.002 \\
        Benchmark 2 & 41.992 & 2.170 & 194.168 & [40, 7, 27] & 4.926 & 2.072 & 7.492 \\
        Benchmark 3 & 47.439 & 1.720 & 208.186 & [35, 20, 27] & 3.723 & 1.370 & 7.882 \\
        Benchmark 4 & 4.474 & 2.047 & 9.482 & [28] & 3.151 & 3.378 & 1.950 \\
        \bottomrule
    \end{tabular}
\end{table}

\added[id=PG]{Both orchestrated variants attain a lower mean chamfer distance than any of the four single-agent benchmarks, on the raw statistics ($3.496$ and $3.634$ against $4.474$ to $47.439$) and with anomalies excluded ($2.181$ and $1.846$ against $2.696$ to $4.926$). The advantage lies in the tail rather than the typical case: one benchmark configuration attains the lowest median of any run in the table ($1.384$), and holds that position once anomalies are removed, so on a part that goes well a single agent with the same tools is competitive. What separates the two is how badly things go when they go wrong. Two of the four benchmark configurations carry standard deviations of $194.168$ and $208.186$, an order of magnitude beyond anything the orchestrated pipeline produces, because a single agent that chooses when to compile and when to render can skip those checks and compound an unnoticed error across the remaining transitions. Holding the orchestration fixed and changing only the model (Variant~B against Variant~F) moves the summary statistics far less than removing the orchestration does, which suggests the structure contributes more here than the choice between the two Gemini versions.}

\todo[inline]{\cref{tab:benchmark}: (1) label which model / transition-batching combination each of Benchmarks 1--4 corresponds to, so the $2\times2$ design is readable. (2) The Benchmark~4 raw statistics (4.474 / 2.047 / 9.482) are identical to Variant~C in \cref{tab:ablation}; verify this is not a copy-paste. (3) Kojo's write-up ends here --- no prose interpreting the benchmark results yet.}

\section{Limitations and Ongoing Work}

This is a preliminary report, and the evaluation above characterizes the CAD Modifier rather than the complete {DFM-Redesign pipeline}. We record here what the present evidence does and does not support.

\paragraph{Statistical strength.} Each variant was run once over the 46-part set, and the comparisons are reported as means, medians and win counts without significance testing. Given the run-to-run variability documented above, differences between neighbouring variants should be read as indicative rather than established, and the ordering of the closest variants may not survive replication. The outliers excluded from the ``without anomalies'' columns were identified per variant by inspection rather than by a pre-declared criterion, so those columns summarize differently sized subsets and are not directly comparable across rows. Repeated runs under a uniform, pre-declared outlier rule, with paired significance tests, are in progress.

\paragraph{Manufacturability improvement is not yet quantified.} We have not measured the DFM gain of a redesigned part over its original. The intended measures are estimated machining cost and time deltas, the number of setups and tool changes, and the removal of hard-to-machine features such as sharp internal vertical corners and fillets requiring ball-nose tooling.

\paragraph{Design-intent preservation lacks an operational definition.} The claim that redesigns preserve intent currently rests on visual inspection. Substantiating it requires a pre-declared metric --- geometric similarity, retained functional features, and dimensional deviation within a stated tolerance --- rather than a subjective judgement.

\paragraph{Reachable complexity.} The pump casing demonstrates that chained verified transitions reach parts that one-shot generation does not, but we have not yet plotted transition success rate against part complexity. That sweep is the direct evidence for the paper's central methodological claim and is the most important outstanding experiment.

\paragraph{Evaluators and baselines.} Fitness-for-purpose has not been judged by human or AI evaluators; doing so calls for a judge from a different model family than the system under test, to avoid Gemini assessing Gemini, with inter-rater agreement reported. Beyond the single-agent benchmarks reported above, comparisons against manual DFM review, an established DFM tool such as DFMXpress~\citep{dfmxpress} or DFMPro~\citep{dfmpro}, and prior piecewise or edit-based CAD generation methods~\citep{zhang2025flexcad,xu2024cadmllm,xu2024brepgen} remain outstanding. The latter are not DFM systems, so the comparison there is on edit controllability, fidelity to the instructed change, and reachable complexity rather than on manufacturability outcomes.

\paragraph{Benchmark configurations.} The four benchmark runs in \cref{tab:benchmark} are not yet individually labeled by model and transition-batching strategy, and one row is pending re-verification against its source data. That comparison should therefore be read at the level of orchestrated pipeline against single agent, not between individual benchmark rows.

\paragraph{Dataset release.} We intend to release the 46-part evaluation set --- multi-view image sets, STEP files, and the manually-reviewed transition descriptions --- together with the complexity spread of its parts.

\section{Conclusion}

We presented the {DFM-Redesign pipeline}, a training-free multi-agent system that reviews a CAD part for manufacturability and redesigns it through a sequence of individually verified design transitions. Applying one edit at a time, and verifying each against the intended change --- first by compilation, then by visual inspection of multi-view renderings --- lets the system compound controlled edits without the fidelity loss that limits one-shot X-to-CAD generation. On a 46-part benchmark scored by chamfer distance to reference geometries, the CAD Modifier attains a lower mean chamfer distance than chain-of-thought single-agent baselines equipped with the same tooling, and our ablations isolate the contribution of the visual review loop and of the pre-transition captioning step. Because the manufacturing knowledge comes from a general-purpose multimodal model rather than from fine-tuning, the approach retargets directly onto stronger models as they arrive. The end-to-end evaluation of the review-and-redesign loop is ongoing and will appear in a subsequent revision.

\section*{Acknowledgments}

This material is based upon work supported by the National Science Foundation under Award No.~2328032, \emph{FMSG: Cyber: Learning Foundation Models for Manufacturing Design Automation}.

\bibliographystyle{icml2026}
\bibliography{references}

\end{document}